\documentclass[10pt]{amsart}

\usepackage{amsthm,amsmath,amssymb}
\usepackage[dvips]{graphicx}

\newcommand{\F}{\mathbb{F}}

\newcommand{\R}{\mathbb{R}}

\newcommand{\A}{\mathbb{A}}

\newcommand{\ga}{\alpha}
\newcommand{\gb}{\beta}

\renewcommand{\phi}{\varphi}

\newcommand{\calL}{\mathcal{L}}
\newcommand{\calO}{\mathcal{O}}

\newcommand{\set}[1]{\{#1\}}
\newcommand{\ideal}[1]{\langle #1\rangle}

\DeclareMathOperator{\ev}{ev}

\DeclareMathOperator{\LT}{lt}
\DeclareMathOperator{\LC}{lc}
\DeclareMathOperator{\ind}{ind}

\newcommand{\zdeg}{\text{$z$-$\deg$}}

\newtheorem{thm}{Theorem}
\newtheorem{prop}[thm]{Proposition}
\newtheorem{lem}[thm]{Lemma}

\theoremstyle{definition}
\newtheorem*{exmp}{Example}
\newtheorem*{algI}{Algorithm I}
\newtheorem*{algG}{Algorithm G}

\begin{document}

\title[List Decoding of Hermitian Codes using Gr\"obner Bases]{List Decoding of Hermitian Codes \\ using Gr\"obner Bases}

\author{Kwankyu~Lee}
\address{Korea Institute of Advanced Study, Seoul, Korea}
\email{kwankyu@kias.re.kr}
\thanks{The first author was supported by the Korea Research Foundation Grant funded by the Korea Government (MOEHRD, Basic Research Promotion Fund) (KRF-2005-214-C00009).}

\author{Michael E.~O'Sullivan} 
\address{Department of Mathematics and Statistics, San Diego State University, San Diego, USA}
\email{mosulliv@math.sdsu.edu}

\begin{abstract}
List decoding of Hermitian codes is reformulated to allow an efficient and simple algorithm for the interpolation step. The algorithm is developed using the theory of Gr\"obner bases of modules. The computational complexity of the algorithm seems comparable to previously known algorithms achieving the same task, and the algorithm is better suited for hardware implementation.
\end{abstract}

\maketitle

\section{Introduction}

Following Sudan's idea of list decoding of Reed-Solomon codes \cite{sudan1997}, Shokrollahi and Wasserman \cite{shokrollahi1999} presented the first form of list decoding of algebraic geometry codes. Soon afterward, Guruswami and Sudan \cite{guruswami1999} added the notion of multiplicities to Shokrollahi and Wasserman's formulation, improving significantly the capability of list decoding. By these works, the current form of list decoding of algebraic geometry codes, consisting of an interpolation step and a root-finding step, was established.

Subsequently, many efforts followed to develop practical algorithms for the interpolation step and the root-finding step. H\o holdt and Nielsen \cite{hoeholdt1999} worked out explicitly an interpolation algorithm and a factorization algorithm specifically for Hermitian codes. Augot and Pecquet \cite{augot2000}, Gao and Shokrollahi \cite{gao2000}, and Wu and Siegel \cite{wu2001} presented efficient factorization or root-finding algorithms over function fields. Sakata \cite{sakata2001} presented a fast interpolation method using the well-known Berlekamp-Massey-Sakata algorithm. Olshevsky and Shokrollahi \cite{olshevsky2003} derived fast interpolation algorithms using the concept of displacement rank of structured matrices. 

Hermitian codes have been the most prominent example of algebraic geometry codes, and a serious competitor of Reed-Solomon codes. First of all, they are significantly longer than Reed-Solomon codes for a fixed alphabet size, and they have good dimension and minimum distance properties. They also possess a rich algebraic and geometric structure that yields efficient encoding and decoding algorithms. New developments on decoding algorithms were often applied to Hermitian codes foremost, and an idea successful with Hermitian codes is likely to be extended for a general class of algebraic geometry codes.

The aim of this paper is to extend the results of Lee and O'Sullivan \cite{kwankyu2006} for Hermitian codes. This is a natural but, we think,  nontrivial task. We needed to reformulate list decoding of Hermitian codes in the language of commutative algebra and Gr\"obner bases. An advantage of the new formulation is to eliminate the computation of the ``increasing zero bases'' of a linear space as in \cite{hoeholdt1999}. The new formulation allows us to present a simple and efficient algorithm for the interpolation step using Gr\"obner bases of modules. The algorithm is a natural adaptation to Hermitian codes of the algorithm for Reed-Solomon codes developed in \cite{kwankyu2006}. 

In Section 2, we review basic properties of Hermitian curves and codes. Fulton \cite{fulton1969}, Stichtenoth \cite{stichtenoth1993}, and Pretzel \cite{pretzel1998} are our basic references for further information. In later sections, a basic understanding of Gr\"obner bases is assumed. For an introduction to the theory, see Cox et al.~\cite{cox1997,cox1997a}. In Section 3, we formulate list decoding of Hermitian codes. In Section 4, we decribe a method to find an optimal interpolation polynomial, namely the $Q$-polynomial. In Section 5, an efficient algorithm for the interpolation step is presented. In Section 6, some upper bounds for the $Q$-polynomial are given. In the appendix, we present an algorithm computing a Gr\"obner basis for a module with a special set of generators, with respect to a special weight monomial order. It is a slight abstraction of Algorithm G for list decoding of Reed-Solomon codes presented in \cite{kwankyu2006}, and applicable for Hermitian codes as well.

\section{Codes on Hermitian curves}

Let $\F$ denote a finite field with $q^2$ elements. Let $H\subset\A^2_\F$ be the Hermitian plane curve defined by the absolutely irreducible polynomial $X^{q+1}-Y^q-Y$ over $\F$. The coordinate ring of $H$ is the integral domain
\[
	R=\F[X,Y]/\ideal{X^{q+1}-Y^q-Y}.
\]
The function field of $H$ is the quotient field $K$ of $R$. Let $x$ and $y$ denote the residue classes of $X$ and $Y$ in $R$, respectively. So $x^{q+1}-y^q-y=0$, and $R=\F[x,y]$. 

There are $q^3$ rational points on $H$, which are enumerated as $P_1,P_2,\dots,P_n$ with $n=q^3$. The projective closure of $H$ is a nonsingular curve with a unique rational point $P_\infty$ at infinity. The functions $x$ and $y$ on $H$ have poles at $P_\infty$ of orders $q$ and $q+1$, respectively. The genus of $H$ is given by $g=q(q-1)/2$. 

The linear space $\calL(uP_\infty)$ for $u\ge 0$ has a basis consisting of $x^iy^j$ for $0\le i$, $0\le j\le q-1$, and $qi+(q+1)j\le u$. Moreover
\[
	R=\bigcup_{u=0}^\infty\calL(uP_\infty)=\bigoplus_{\substack{0\le i \\ 0\le j\le q-1}}\!\!\F\cdot x^iy^j.
\]

Recall that the Hamming space $\F^n$ is equpped with the Hamming distance $d$. Let $P_i=(\ga_i,\gb_i)$ for $1\le i\le n$. The evaluation map $\ev:R\to\F^n$ defined by
\[
	\phi\mapsto(\phi(P_1),\phi(P_2),\dots,\phi(P_n))
\]
is a linear map over $\F$. We now fix a positive integer $u$. Hermitian code $C_u$ is defined to be the linear code given as the image of $\calL(uP_\infty)$ by the evaluation map. If $u<n$, then $\ev$ is injective on $\calL(uP_\infty)$, and the dimension of $C_u$ is equal to $\dim_\F(\calL(uP_\infty))$, which is $u+1-g$ for $u\ge 2g$ by the Riemann-Roch theorem. Note also that the minimum distance of $C_u$ is at least $n-u$.

Let $k$ denote the dimension of $C_u$. For encoding, fix a basis of $\calL(uP_\infty)$, say $\set{\phi_1,\phi_2,\dots,\phi_k}$. Then a message $\omega=(\omega_1,\omega_2,\dots,\omega_k)\in\F^k$ is encoded to the codeword 
\[
	c=\ev(\mu),\quad
	\mu=\sum_{i=1}^k\omega_i\phi_i.
\]
We call $\mu$ the message function corresponding to the codeword $c$.

\begin{exmp}
Let $q=2$. We consider the Hermitian curve $H$ defined by $X^3+Y^2+Y$ over $\F_4=\set{0,\ga,\ga^2,\ga^3}$. There are $8$ rational points on $H$, 
\[
	(0,0),(0,1),(1,\ga),(1,\ga^2),(\ga,\ga),(\ga,\ga^2),(\ga^2,\ga),(\ga^2,\ga^2).
\]
Let $u=4$. The linear space $\calL(4P_\infty)$ has a basis $\set{1,x,y,x^2}$. Hermitian code $C_4$ is an $[8,4,4]$ linear code over $\F_4$. Our message is $\omega=(\ga^2,\ga^2,0,\ga^2)$, which is encoded to the codeword
\[
	\ev(\mu)=(\ga^2,\ga^2,\ga^2,\ga^2,0,0,0,0),
\]
where $\mu=\ga^2+\ga^2x+\ga^2x^2$. We will continue this example throughout.
\end{exmp}

Define
\[
	H_i=-\frac{(X^{q^2}-X)(Y^q+Y-\gb_i^q-\gb_i)}{(X-\ga_i)(Y-\gb_i)}\in\F[X,Y]
\]
and let $h_i$ denote the residue class of $H_i$ in $R$ for $1\le i\le n$. 

For $v=(v_1,v_2,\dots,v_n)\in\F^n$, define $H_v=\sum_{i=1}^n v_iH_i$, and let $h_v$ denote the residue class of $H_v$ in $R$. That is, 
\[
	h_v=\sum_{i=1}^n v_ih_i.
\]

\begin{lem}
$h_i(P_j)=1$ if $j=i$, and $0$ otherwise. So $\ev(h_v)=v$ for $v\in\F^n$. 
\end{lem}

\begin{proof}
Recall that
\[
	\prod_{a\in\F}(X-a)=X^{q^2}-X,
	\quad
	\prod_{\substack{b\in\F\\b^q+b=\gb^q+\gb}}(Y-b)=Y^q+Y-\gb^q-\gb
\]
for any $\gb\in\F$. For $1\le i\le n$, define
\[
\tilde{H}_i
	=\prod_{\substack{a\in\F \\ a\neq\ga_i}}(X-a)\prod_{\substack{b\in\F,\,b \neq \gb_i \\ b^q+b=\gb_i^q+\gb_i=\ga_i^{q+1}}}(Y-b)
	=\frac{(X^{q^2}-X)(Y^q+Y-\gb_i^q-\gb_i)}{(X-\ga_i)(Y-\gb_i)}.
\]
It is immediate from the definition that $\tilde{H}_i(\ga_j,\gb_j)=0$ for $j\neq i$. Taking partial derivatives with respect to $X$ and $Y$ of both sides of the equation
\[
	(X-\ga_i)(Y-\gb_i)\tilde{H}_i=(X^{q^2}-X)(Y^q+Y-\gb_i^q-\gb_i)
\]
and substituting $X$ and $Y$ with $\ga_i$ and $\gb_i$, we see that $\tilde{H}_i(\ga_i,\gb_i)=-1$. As $h_i$ is the residue class of $-\tilde{H}_i$ in $R$, the assertion follows.
\end{proof}

\begin{exmp}[continued]
The functions $h_i$ are as follows:
\begin{align*}
	h_1&=(x^3+1)y+x^3+1,\\
	h_2&=(x^3+1)y,\\
	h_3&=(x^3+x^2+x)y+\ga^2x^3+\ga^2x^2+\ga^2x,\\
	h_4&=(x^3+x^2+x)y+\ga x^3+\ga x^2+\ga x,\\
 	h_5&=(x^3+\ga x^2+\ga^2x)y+\ga^2x^3+x^2+\ga x,\\
	h_6&=(x^3+\ga x^2+\ga^2x)y+\ga x^3+\ga^2x^2+x,\\
	h_7&=(x^3+\ga^2x^2+\ga x)y+\ga^2x^3+\ga x^2+x,\\
	h_8&=(x^3+\ga^2x^2+\ga x)y+\ga x^3+x^2+\ga^2x.
\end{align*}
\end{exmp}

Lastly, define $\eta$ to be the residue class of $\prod_{a\in\F}(X-a)=X^{q^2}-X$ in $R$. So $\eta=x^{q^2}-x$.

\section{List Decoding of Hermitian Codes}

We prove some lemmas required for a fundamental theorem, Theorem \ref{thmakwe}, of list decoding of Hermitian codes. First, note that the surface $S=H\times\A^1_\F$ has coordinate ring
\[
	R[z]=\F[X,Y,Z]/\ideal{X^{q+1}-Y^q-Y},
\]
where $z$ denotes the residue class of $Z$ in the quotient ring.

\begin{lem}\label{lemabcd}
Let $m$ be a positive integer. Let $v$ be a vector in $\F^n$. Then 
\[
	\dim_{\F} R[z]/\ideal{z-h_v,\eta}^m=n\binom{m+1}{2}.
\]
Let $\mu\in R$ with $t=d(v,\ev(\mu))$. Then
\[
	\dim_\F R[z]/(\ideal{z-h_v,\eta}^m+\ideal{z-\mu})=m(n-t).
\]
\end{lem}

\begin{proof}
Let $k$ be an algebraic closure of $\F$. We consider the ideal 
\[
	I=\ideal{X^{q+1}-Y^q-Y}+\ideal{Z-H_v,X^{q^2}-X}^m
\]
of $k[X,Y,Z]$. We claim that the zero set $V(I)$ of $I$ is $\set{(\ga_i,\gb_i,v_i)\mid 1\le i\le n}$. As one inclusion is easily verified, we show that every $(a,b,c)\in V(I)$ equals $(\ga_i,\gb_i,v_i)$ for some $1\le i\le n$. Suppose $(a,b,c)\in V(I)$. Since $(X^{q^2}-X)^m\in I$, we have $a^{q^2}-a=0$.  So $a\in\F$. We also have $a^{q+1}-b^q-b=0$. Taking the $q$-th power of the equation, we get $a^{q+1}-b^{q^2}-b^q=0$. Subtracting the second equation from the first, we get $b^{q^2}-b=0$. Therefore $b\in\F$. Thus $(a,b)$ must be one of the rational points on the Hermitian curve. Let $(a,b)=P_i$ for some $1\le i\le n$. As $(Z-H_v)^m\in I$, we see $c=v_i$. The claim is now proved.

As $V(I)$ is finite, we have a natural isomorphism (see Theorem 2.2 in Chapter 4 of \cite{cox1997a})
\[
	k[X,Y,Z]/I\cong\bigoplus_{i=1}^n \calO_{(\ga_i,\gb_i,v_i)}/I\calO_{(\ga_i,\gb_i,v_i)}
\]
where $\calO_{(\ga_i,\gb_i,v_i)}$ denotes the local ring $k[X,Y,Z]_{\ideal{X-\ga_i,Y-\gb_i,Z-v_i}}$. Now fix $i$. The automorphism of $k[X,Y,Z]$ given by $(X,Y,Z)\mapsto(X+\ga_i,Y+\gb_i,Z+v_i)$ induces the isomorphism
\[
	\calO_{(\ga_i,\gb_i,v_i)}/I\calO_{(\ga_i,\gb_i,v_i)}\cong \calO/I'\calO
\]
where $\calO=k[X,Y,Z]_{\ideal{X,Y,Z}}$, and $I'$ is the ideal 
\[
	I'=\ideal{X^{q+1}+(\ga_i^q+\ga_i)X-Y^q-Y}+\ideal{Z+AX+BY,X^{q^2}-X}^m
\]
for some $A,B\in k[X,Y]$. As $V(I')$ is finite and contains the origin, we have
\[
	\dim_k\calO/I'\calO=\dim_kk[[X,Y,Z]]/I'k[[X,Y,Z]].
\]
In $k[[X,Y,Z]]$, we can write by the Weierstrass Preparation Theorem, 
\[
	Y^q+Y-X^{q+1}-(\ga_i^q+\ga_i)X=(Y-XP)U
\]
for some $P\in k[[X]]$ and a unit $U$ of $k[[X,Y]]$. As an ideal of $k[[X,Y,Z]]$,
\[
\begin{split}
	I'k[[X,Y,Z]]
		&=\ideal{Y-XP}+\ideal{Z+(A+BP)X,X^{q^2}-X}^m\\
		&=\ideal{Y-XP}+\ideal{Z+(A+BP)X,X}^m\\
		&=\ideal{Y-XP}+\ideal{Z,X}^m.
\end{split}
\]
So $k[[X,Y,Z]]/I'k[[X,Y,Z]]\cong k[X,Z]/\ideal{X,Z}^m$.
Since this is true for all $i$, 
\[
	\dim_kk[X,Y,Z]/I=n\dim_kk[X,Z]/(X,Z)^m=n\binom{m+1}{2}.
\]
The first assertion of the lemma now follows since
\[
	\dim_\F R[z]/\ideal{z-h_v,\eta}^m=\dim_k k[X,Y,Z]/I
\]
as $I$ is an ideal generated by polynomials defined over $\F$.

The second assertion is proved similarly. So we will be brief, omitting repeated details. Let 
\[
	J=\ideal{X^{q+1}-Y^q-Y}+\ideal{Z-H_v,X^{q^2}-X}^m+\ideal{Z-M},
\]
where $M\in\F[X,Y]$ is such that $\mu$ is the residue class of $M$ in $R$. Let $\ev(\mu)=(c_1,c_2,\dots,c_n)$.
Then $V(J)=\set{(\ga_i,\gb_i,v_i)\mid v_i=c_i, 1\le i\le n}$. We have a natural isomorphism
\[
	k[X,Y,Z]/J\cong\bigoplus_{v_i=c_i} \calO_{(\ga_i,\gb_i,v_i)}/J\calO_{(\ga_i,\gb_i,v_i)}.
\]
Fix $i$ with $v_i=c_i$. Then $\calO_{(\ga_i,\gb_i,v_i)}/J\calO_{(\ga_i,\gb_i,v_i)}$ is isomorphic to $\calO/J'\calO$, where 
\[
\begin{split}
	J'	&=\ideal{X^{q+1}+(\ga_i^q+\ga_i)X-Y^q-Y}+\ideal{Z+AX+BY,X^{q^2}-X}^m\\
		&\quad+\ideal{Z+CX+DY}
\end{split}
\]
for some $A,B,C,D\in k[X,Y]$. Again 
\[
	\dim_k\calO/J'\calO=\dim_kk[[X,Y,Z]]/J'k[[X,Y,Z]],
\]
but now
\[
	J'k[[X,Y,Z]]=\ideal{Y-XP}+\ideal{X}^m+\ideal{Z+SX}
\]
for some $P,S\in k[[X]]$. This gives an isomorphism
\[
	k[[X,Y,Z]]/J'k[[X,Y,Z]]\cong k[X]/\ideal{X}^m.
\]
Therefore
\[
	\dim_kk[X,Y,Z]/J=\sum_{v_i=c_i}\dim_kk[X]/\ideal{X}^m=m(n-t),
\]
from which the second assertion of the lemma follows.
\end{proof}

\begin{lem}
Let $\psi$ be a nonzero element in $R$. Then 
\[
	\dim_\F(R/\psi)=-v_{P_\infty}(\psi).
\]
\end{lem}

\begin{proof}
Recall that $P_\infty$ is the unique point at infinity of the smooth curve $H$. Therefore 
\[
	R=\bigcap_{P\neq P_\infty}\calO_P.
\]
Consider the homomorphism 
\[
	g: R\longrightarrow\bigoplus_{P\neq P_\infty}\calO_P/\psi\calO_P
\]
which maps $\phi\in R$ to $\overline{\phi}$ in $\calO_P/\psi\calO_P$ for each $P\neq P_\infty$. If $\phi\in\ker(g)$, then $\phi/\psi\in\calO_P$ for $P\neq P_\infty$, which implies $\phi/\psi\in R$, and hence $\phi\in\psi R$. Therefore $\ker(g)=\psi R$. To prove surjectivity, let $S$ be the finite set of points of $H$ at which $v_P(\psi)>0$. Let $(\overline{\chi_P})$ be an element of the direct sum. Then by the Strong Approximation Theorem, there is a $\phi$ in the function field $K$ such that $v_P(\phi-\chi_P)=v_P(\psi)$ for $P\in S$ and $v_P(\phi)\ge 0$ for $P\notin S$ and $P\neq P_\infty$. Then 
\[
	\phi\in\bigcap_{P\neq P_\infty}\calO_P=R,\quad\text{and}\quad	\phi\equiv\chi_P\mod \psi O_P
\]
for $P\neq P_\infty$. This shows that $g$ is surjective. 

Hence we have a natural isomophism 
\[
	R/\psi\overset{\cong}{\longrightarrow}\bigoplus_{P\neq P_\infty}\calO_P/\psi\calO_P,
\]
which implies
\[
	\dim_\F(R/\psi)=\sum_{P\neq P_\infty}\dim_\F(\calO_P/\psi\calO_P)
		=\sum_{P\neq P_\infty}v_P(\psi)
		=-v_{P_\infty}(\psi).
\]
\end{proof}

We introduce two notations. For $f\in R[z]$, the $u$-weighted degree of $f$ is defined to be
\[
	\deg_u(f)=\max_{0\le i\le a}(-v_{P_\infty}(\psi_i)+ui)
\]
if $f=\psi_az^a+\dots+\psi_1z+\psi_0$. For $f\in R[z]$ and $\phi\in R$, we denote by $f(\phi)$ the element in $R$ that is obtained by substituting $z$ with $\phi$ in $f$. Observe that if $\phi\in\calL(uP_\infty)$, then $-v_{P_\infty}(f(\phi))\le\deg_u(f)$.

Now we are ready to present a fundamental theorem upon which list decoding of Hermitian codes is based.
Suppose that some codeword of $C_u$ was sent through a noisy channel. Let $v$ denote the vector in $\F^n$ that was received by hard-decision on the channel output. Fix a positive integer $m$, called the multiplicity parameter. Define
\[
	I_{v,m}=\ideal{z-h_v,\eta}^m,
\]
an ideal of the integral domain $R[z]$.

\begin{thm}\label{thmakwe}
Suppose $f\in I_{v,m}$ is nonzero. Let $w=\deg_u(f)$. If $c$ is a codeword of $C_u$ satisfying
\[
	d(v,c)<n-w/m,
\]
then $f(\mu)=0$, where $\mu$ is the message function corresponding to $c$.
\end{thm}

\begin{proof}
Let $t=d(v,c)$. Assume $f(\mu)$ is not zero in $R$. Then
\[
\begin{split}
	w=\deg_u(f)	&\ge-v_{P_\infty}(f(\mu))\\
				&=\dim_\F(R/f(\mu)) \\
				&=\dim_\F(R[z]/\ideal{f,z-\mu})\\
				&\ge\dim_\F(R[z]/(\ideal{z-h_v,\eta}^m+\ideal{z-\mu})=m(n-t).
\end{split}
\]
Therefore if $m(n-t)>w$, we must have $f(\mu)=0$. 
\end{proof}

The first step of list decoding of Hermitian codes is to construct a nonzero $f$ in $I_{v,m}$. The second step is to find roots of $f$ over $R$, and output the list of message functions corresponding to codewords of $C_u$. To maximize the possibility that the list contains the original message function corresponding to the sent codeword, $f$ should be chosen such that the $u$-weighted degree of $f$ is minimized, according to Theorem \ref{thmakwe}.

\section{Using Gr\" obner Bases of Modules}

We call the elements in the set 
\[
	\Omega=\set{x^iy^jz^k\mid 0\le i, 0\le j\le q-1, 0\le k}
\]
monomials of $R[z]$. Recall that every element of $R[z]$ can be written as a unique linear combination over $\F$ of monomials of $R[z]$. Note that 
\[
	\deg_u(x^iy^jz^k)=qi+(q+1)j+uk.
\]
For two monomials $x^{i_1}y^{j_1}z^{k_1}$, $x^{i_2}y^{j_2}z^{k_2}$ in $\Omega$, we declare 
\[
	x^{i_1}y^{j_1}z^{k_1}>_u x^{i_2}y^{j_2}z^{k_2}
\] 
if $\deg_u(x^{i_1}y^{j_1}z^{k_1})>\deg_u(x^{i_2}y^{j_2}z^{k_2})$ or $k_1>k_2$ when tied. It is easy to verify that $>_u$ is a total order on $\Omega$. Notions such as the leading term and the leading coefficient of $f\in R[z]$ are defined in the usual way. For $f\in R[z]$, the $z$-degree of $f$, written $\zdeg(f)$, is the degree of $f$ as a polynomial in $z$ over $R$.

Now we define the $Q$-polynomial of $I_{v,m}$ as the unique, up to a constant multiple, element in $I_{v,m}$ with the smallest leading term with respect to $>_u$. By the definition, the $Q$-polynomial is an element of $I_{v,m}$ with the smallest $u$-weighted degree, and moreover it has the smallest $z$-degree among such elements. Therefore we may say that the $Q$-polynomial is an optimal choice for the interpolation step of list decoding, and that the goal of the interpolation step is to find the $Q$-polynomial efficiently. We now present our strategy for this task in the following.

Let $Q$ denote the $Q$-polynomial of $I_{v,m}$ from now on. Let $l$ be a positive integer such that $\zdeg(Q)\le l$. We call $l$ the list size parameter. Define
\[
	R[z]_l=\set{f\in R[z]\mid \zdeg(f)\le l}.
\]
Note that $R[z]_l$ is a free module over $R$ of rank $l+1$ with a free basis $1,z,\dots,z^l$. Define $I_{v,m,l}=I_{v,m}\cap R[z]_l$. Clearly $I_{v,m,l}$ is a submodule of $R[z]_l$ over $R$.
\begin{prop}
$I_{v,m,l}$, as a module over $R$, has a set of generators consisting of $G_i$, $0\le i\le l$, where
\[
	G_i=
	\begin{cases}
	(z-h_v)^i\eta^{m-i}&\quad 0\le i\le m,\\
	z^{i-m}(z-h_v)^m&\quad m<i.
	\end{cases}
\]
\end{prop}

\begin{proof}
Recall that $I_{v,m}$ is generated by $G_i$, $0\le i\le m$ as an ideal of $R[z]$. Note that for $0\le i<m$,
\[
	z(z-h_v)^i\eta^{m-i}=\eta(z-h_v)^{i+1}\eta^{m-i-1}+h_v(z-h_v)^i\eta^{m-i}.
\]
Using this equation repeatedly, we may write any $f\in I_{v,m,l}$ as a linear combination of the $G_i$ with coefficients in $R$. Then since $\zdeg(f)\le l$, the coefficient of $G_i$ for $i>l$ in the linear combination must be zero. This completes the proof.
\end{proof}

Observe that the ring $R=\F[x,y]$ is in turn a free module over $\F[x]$ of rank $q$, with a free basis $\set{1,y,\dots,y^{q-1}}$. This can be seen easily from the relation $y^q=-y+x^{q+1}$. So we may view $R[z]_l$ as a free module of rank $q(l+1)$ over $\F[x]$ with a free basis $\set{y^jz^i\mid 0\le i\le l, 0\le j\le q-1}$. The elements of $\Omega\cap\R[z]_l$ will be called monomials of $R[z]_l$. It is clear that the total order $>_u$ is precisely a monomial order on the free module $R[z]_l$ over $\F[x]$.
We also view $I_{v,m,l}$ as a submodule of the free module $R[z]_l$ over $\F[x]$. A set of generators for $I_{v,m,l}$, as a module over $\F[x]$, is
\[
	\set{y^jG_i\mid 0\le i\le l, 0\le j\le q-1}.
\]
It is immediate that the $Q$-polynomial of $I_{v,m}$ is also the element of $I_{v,m,l}$ with the smallest leading term with respect to $>_u$. As a consequence of the definition of Gr\"obner bases, $Q$ occurs as the smallest element in any Gr\"obner basis of the module $I_{v,m,l}$ over $\F[x]$ with respect to $>_u$. Unlike the computation of Gr\"obner bases of ideals, it turns out that the computation of a Gr\"obner basis of the module $I_{v,m,l}$ over $\F[x]$ can be done efficiently. 

\begin{exmp}[continued]
Suppose the received vector is 
\[
	v=(\ga^2,0,0,\ga^2,0,0,0,0).
\]
Our multiplicity parameter is $m=2$. Then
\[
	h_v=\ga^2x^2y+\ga x^3+\ga^2xy+x^2+\ga^2y+x+\ga^2,
\]
and $\eta=x^4+x$. It will turn out that $\zdeg(Q)=2$. So we take $l=2$ as our list size parameter. As a module over $R=\F_4[x,y]$ with $y^2=y+x^3$,
\[
	I_{v,2,2}=\langle z+h_v,\eta \rangle^2=\langle\eta^2,\eta z+\eta h_v,z^2+h_v^2\rangle=\langle G_0,G_1,G_2\rangle,
\]
where
\begin{align*}
	G_0	&=x^8+x^2, \\
	G_1	&=(x^4+x)z+(\ga^2 x^6+\ga^2 x^5+\ga^2 x^4+\ga^2 x^3+\ga^2 x^2+\ga^2 x)y\\
		&\quad+\ga x^7+x^6+x^5+x^4+x^3+x^2+\ga^2 x, \\
	G_2	&=z^2+(\ga x^4+\ga x^2+\ga)y+\ga x^7+\ga^2x^6+\ga x^5+x^4+\ga x^3+x^2+\ga.
\end{align*}
As a module over $\F_4[x]$, $I_{v,2,2}=\langle G_0,yG_0,G_1,yG_1,G_2,yG_2 \rangle$,
where
\begin{align*}
	yG_0&=(x^8+x^2)y,\\
	yG_1&=(x^4+x)yz+(\ga x^7+\ga x^6+\ga x^5+\ga x^4+\ga x^3+\ga x^2)y\\
		&\quad+\ga^2 x^9+\ga^2 x^8+\ga^2 x^7+\ga^2 x^6+\ga^2 x^5+\ga^2 x^4,\\
	yG_2&=yz^2+(\ga x^7+\ga^2 x^6+\ga x^5+\ga^2 x^4+\ga x^3+\ga^2 x^2)y+\ga x^7+\ga x^5+\ga x^3.
\end{align*}
\end{exmp}

\section{An Interpolation Algorithm }

We obtain an interpolation algorithm for Hermitian codes, applying Algorithm G in the appendix to the free module $R[z]_l$ over $\F[x]$ and the set of generators $y^jG_i$ of the submodule $I_{v,m,l}$ of $R[z]_l$ as given in the previous section.

Let $T=\set{(i,j)\mid 0\le i\le l, 0\le j\le q-1}$. Tuples in $T$ are ordered lexicographically. So $(0,0)$ is the first tuple in $T$ and the successor of $(i,j)$ is $(i,j+1)$ if $j<q-1$ or $(i+1,0)$ if $j=q-1$. Thus $\set{y^jz^i\mid (i,j)\in T}$ is a basis for $R[z]_l$ as an $\F[x]$-module and the weight of the basis element $y^jz^i$ is $ui+(q+1)j$. The index of $f\in R[z]_l$ is the largest tuple $(i,j)$ such that the coefficient of $y^jz^i$ is nonzero. So if the leading term, with respect to $>_u$, of $f\in R[z]_l$ is $x^iy^jz^k$, then $\ind(\LT(f))=(k,j)$. Notice that $\ind(y^jG_i)=(i,j)$. 

\begin{algI} 
The algorithm finds the element of $I_{v,m,l}$ with the smallest leading term. Initially set $g_{i,j}\leftarrow y^jG_i$ for $(i,j)\in T$. Let 
\[
	g_{i,j}=\sum_{(i',j')\in T}a_{i,j,i',j'}y^{j'}z^{i'}
\]
during the execution of the algorithm. For $r=(r_1,r_2)$ and $s=(s_1,s_2)$ in $T$, the abbreviation $a_{r,s}$ denotes $a_{r_1,r_2,s_1,s_2}$. 
 
\begin{enumerate}
\item[I1.] Set $r\leftarrow (0,0)$.
\item[I2.] Set $r$ to the successor of $r$. If $r\in T$, then proceed; otherwise go to step I6.
\item[I3.] Set $s\leftarrow \ind(\LT(g_r))$. If $s=r$, then go to step I2.
\item[I4.] Set $d\leftarrow\deg(a_{r,s})-\deg(a_{s,s})$ and  $c\leftarrow\LC(a_{r,s})\LC(a_{s,s})^{-1}$.
\item[I5.] If $d\ge 0$, then set 
\[
	g_r\leftarrow g_r-cx^dg_s.
\]
If $d<0$, then set, storing $g_s$ in a temporary variable,
\[
	g_s\leftarrow g_r,\quad g_r\leftarrow x^{-d}g_r-cg_s.
\]
Go back to step I3.
\item[I6.] Output $g_{i,j}$ with the smallest leading term, and the algorithm terminates.  
\end{enumerate}
\end{algI}

The idea of the algorithm is to update the set of generators until $\ind(\LT(g_r))=r$ for all $r\in T$, in which case the updated set of generators is a Gr\"obner basis of $I_{v,m,l}$ trivially by Buchberger's criterion.

\begin{exmp}[continued]
We demonstrate the algorithm by finding the $Q$-polynomial of $I_{v,2}$. In the following, each column corresponds to an element $(i,j)$ of $T$, ordered from right to left. Each entry is a multiple of $y^jz^i$, and the coefficient polynomial from $\F[x]$ is parenthesized with only the leading term shown.

After initialization, we have
{\small\[
	\begin{array}{crrrrrrrrrrrrrrrrr}
	g_{0,0}=&&&&&&&&&&&(x^8+\cdots)\\
	g_{0,1}=&&&&&&&&&(x^8+\cdots)y&&\\
	g_{1,0}=&&&&&&&(x^4+\cdots)z&+&(\ga^2x^6+\cdots)y&+&(\ga x^7+\cdots)\\
	g_{1,1}=&&&&&(x^4+\cdots)yz&+&&&(\ga x^7+\cdots)y&+&(\ga^2x^9+\cdots)\\
	g_{2,0}=&&&z^2&+&&&&&(\ga x^4+\cdots)y&+&(\ga x^7+\cdots) \\
	g_{2,1}=&yz^2&+&&&&&&&(\ga x^7+\cdots)y&+&(\ga x^7+\cdots) 
	\end{array}
\]}%
For $r=(0,0)$ and $(0,1)$, already $\ind(\LT(g_r))=r$. So $r$ proceeds to $(1,0)$. When $r=(1,0)$, we find $s=(0,1)$ in step I3. Since $d=-2$, we update $g_{1,0}$ and $g_{0,1}$ in the second way in step I5. Then we get
{\small\[
	\begin{array}{crrrrrrrrrrrrrrrrr}
	g_{0,0}=&&&&&&&&&&&(x^8+\cdots)\\
	g_{0,1}=&&&&&&&(x^4+\cdots)z&+&(\ga^2 x^6+\cdots)y&+&(\ga x^7+\cdots)\\
	g_{1,0}=&&&&&&&(x^6+\cdots)z&+&(\ga^2 x^7+\cdots)y&+&(\ga x^9+\cdots)\\
	g_{1,1}=&&&&&(x^4+\cdots)yz&+&&&(\ga x^7+\cdots)y&+&(\ga^2x^9+\cdots)\\
	g_{2,0}=&&&z^2&+&&&&&(\ga x^4+\cdots)y&+&(\ga x^7+\cdots) \\
	g_{2,1}=&yz^2&+&&&&&&&(\ga x^7+\cdots)y&+&(\ga x^7+\cdots) 
	\end{array}
\]}%
Now we find $s=(0,0)$ in step I3. Since $d=2$, this time $g_{1,0}$ and $g_{0,1}$ are updated in the first way in step I5. Then we get
{\small\[
	\begin{array}{crrrrrrrrrrrrrrrrr}
	g_{0,0}=&&&&&&&&&&&(x^8+\cdots)\\
	g_{0,1}=&&&&&&&(x^4+\cdots)z&+&(\ga^2 x^6+\cdots)y&+&(\ga x^7+\cdots)\\
	g_{1,0}=&&&&&&&(x^6+\cdots)z&+&(\ga^2 x^7+\cdots)y&+&(x^8+\cdots)\\
	g_{1,1}=&&&&&(x^4+\cdots)yz&+&&&(\ga x^7+\cdots)y&+&(\ga^2x^9+\cdots)\\
	g_{2,0}=&&&z^2&+&&&&&(\ga x^4+\cdots)y&+&(\ga x^7+\cdots) \\
	g_{2,1}=&yz^2&+&&&&&&&(\ga x^7+\cdots)y&+&(\ga x^7+\cdots) 
	\end{array}
\]}%
Now we find $s=(0,1)$ in step I3. Since $d=-1$, we update $g_{1,0}$ and $g_{0,1}$ once again in the second way in step I5. Then we get
{\small\[
	\begin{array}{crrrrrrrrrrrrrrrrr}
	g_{0,0}=&&&&&&&&&&&(x^8+\cdots)\\
	g_{0,1}=&&&&&&&(x^4+\cdots)z&+&(\ga^2 x^6+\cdots)y&+&(\ga x^7+\cdots)\\
	g_{1,0}=&&&&&&&(x^6+\cdots)z&+&&&(\ga^2 x^8+\cdots)\\
	g_{1,1}=&&&&&(x^4+\cdots)yz&+&&&(\ga x^7+\cdots)y&+&(\ga^2x^9+\cdots)\\
	g_{2,0}=&&&z^2&+&&&&&(\ga x^4+\cdots)y&+&(\ga x^7+\cdots) \\
	g_{2,1}=&yz^2&+&&&&&&&(\ga x^7+\cdots)y&+&(\ga x^7+\cdots) 
	\end{array}
\]}%
Finally we find $s=(1,0)=r$ in step I3. That is, $\ind(\LT(g_r))=r$ for $r=(1,0)$. So $r$ is set to the next element in $T$ in step I2. The algorithm proceeds in this way until $\ind(\LT(g_r))=r$ for all $r\in T$. When we reach step I6, we have the following Gr\"obner basis of $I_{v,2,2}$:
{\small\begin{gather*}
	\begin{array}{crrrrrrrrrrrrrrrrr}
	g_{0,0}=&&&z^2&+&\qquad&\qquad&\qquad&\qquad&\qquad&\qquad&\\
	g_{0,1}=&&&(x+\cdots)z^2&+&&&&&&&\\
	g_{1,0}=&yz^2&+&(\ga^2x^2+\cdots)z^2&+&&&&&&&\\
	g_{1,1}=&&&&&&&&&&&\\
	g_{2,0}=&&&(x^2+\cdots)z^2&+&&&&&&&\\
	g_{2,1}=&(x+\cdots)yz^2&+&(\ga x^2+\cdots)z^2&+&&&&&&&
	\end{array}\\
	\begin{array}{crrrrrrrrrrrrrrrrr}
	&&&&&(\ga x^4+\cdots)y&+&(\ga x^7+\cdots)\\
	&&&&&(\ga x^5+\cdots)y&+&(\ga^2 x^6+\cdots)\\
	&&&(\ga^2x^5+\cdots)z&+&(\ga^2 x^5+\cdots)y&+&(x^7+\cdots)\\
	&(x^4+\cdots)yz&+&(\ga^2x^5+\cdots)z&&&&\\
	&&&(\ga^2x^4+\cdots)z&&&&\\
	&&&&&(\ga^2 x^5+\cdots)y&+&(x^6+\cdots)
	\end{array}
\end{gather*}}%
(Here the output for each $g_{i,j}$ is broken into two lines.) Comparing the leading terms in step I6, we find that $g_{2,0}$ is the smallest among the generators. Therefore the algorithm output
\[
	Q=(x^2+x)z^2+(\ga^2x^4+\ga^2x)z,
\]
which has factorization
\[
	Q=(x^2+x)z(z+\ga^2x^2+\ga^2x+\ga^2).
\]
Hence a root-finding algorithm will output the list of roots
\[
	0,\quad\ga^2x^2+\ga^2x+\ga^2,
\]
the second of which is the message function corresponding to the original codeword sent through the channel.
\end{exmp}

\begin{prop}
Aside from the computation of the initial set of generators, an execution of Algorithm I requires $O(n^{8/3}m^2l^3)$ multiplication operations in $\F$.
\end{prop}

\begin{proof}
We rely on Proposition \ref{propxaaa} in the appendix. Note that
\[
	y^jG_i=\sum_{k=0}^i\binom{i}{k}(-1)^{i-k}\eta^{m-i}h_v^{i-k}y^jz^k.
\]
for $0\le i\le m$, $0\le j\le q-1$. Since 
\[
	\deg_u(\eta^{m-i}h_v^{i-k}y^jz^k)\le (m-i)q^3+(i-k)(q^3+q^2-q-1)+j(q+1)+ku,
\]
we see
\[
	\deg_u(y^jG_i)\le m(q^3+q^2-q-1)+q^2-1
\]
and 
\[
	\deg_u(\eta^{m-i}y^jz^i)=(m-i)q^3+j(q+1)+iu\ge mu.
\]
Hence, according to Proposition \ref{propxaaa}, an execution of the algorithm requires
\[
	O((q^3+q^2-q-1)q^{-1}(q^3+q^2-q-1-u)q^3m^2l^3)=O(q^8m^2l^3)
\]
multiplication operations in $\F$.
\end{proof}

\section{Upper bounds for the $Q$-polynomial}

We obtain simple upper bounds on the $u$-weighted degree and the $z$-degree of the $Q$-polynomial of $I_{v,m}$. The $u$-weighted degree of $Q$ determines the number of errors that the list decoder can correct. The $z$-degree of $Q$ is used to set the list size parameter for the list decoder.

\begin{prop}\label{propdkwd}
If $I\subset\Omega$ is a finite set of monomials of $R[z]$ such that 
\[
	|I|>n\binom{m+1}{2},
\]
then there is a set of coefficients $c_\phi\in\F$ such that
\[
	0\neq \sum_{\phi\in I}c_\phi\phi\in I_{v,m}.
\]
\end{prop}

\begin{proof}
The first assertion of Lemma \ref{lemabcd} implies that monomials in $I$ are linearly dependent over $\F$ in $R[z]/I_{v,m}$. On the other hand, they are linearly independent over $\F$ in $R[z]$. This completes the proof. 
\end{proof}

In a table, we arrange monomials of $R[z]$ such that the monomials in the same column have the same $u$-weighted degree and the monomials in the same row have the same $z$-degree. Let weighted degrees increase from left to right and $z$-degrees from bottom to top.

\begin{exmp}[continued]
Recall that $q=2$, $u=4$. So $\deg_u(x^iy^jz^k)=2i+3j+4k$.
\[
\begin{array}{c|*{4}{c}|*{4}{c}|*{4}{c}|*{4}{c}|*{3}{c}}
	3& &    & & &&&   &  &   &        &    &    &z^3 &\bigcirc&\cdots\\
	2& &    & & &&&   &  &z^2&\bigcirc&xz^2&yz^2&x^2z^2&xyz&\cdots\\
	1& &    & & &z&\bigcirc&xz &yz&x^2z&xyz&x^3z&x^2yz&x^4z&x^3yz&\cdots\\
	0&1&\bigcirc&x&y&x^2&xy&x^3&x^2y&x^4&x^3y&x^5&x^4y&x^6&x^5y&\cdots\\\hline
	&0&1&2&3&4&5&6&7&8&9&10&11&12&13
\end{array}
\]
The symbol $\bigcirc$ indicates that there is no monomial for the position.
\end{exmp}

The table of monomials of $R[z]$ suggests the following formula. Let $G(i)=0$ if $i$ is a Weierstrass gap at $P_\infty$, and $1$ otherwise. Note that $G(i)=1$ for $i\ge 2g$. The number of monomials with $u$-weighted degree $i$ is 
\[
	C(i)=\sum_{j=0}^{\lfloor i/u\rfloor}G(i-uj).
\]
Let $w$ be the smallest integer such that
\[
	N=n\binom{m+1}{2}+1\le\sum_{i=0}^wC(i).
\]
Let $l=\lfloor w/u\rfloor$. Then the $u$-weighted degrees and the $z$-degrees of monomials up to the $N$th monomial are not greater than $w$ and $l$, respectively. Now Proposition \ref{propdkwd} implies $\deg_u(Q)\le w$ and $\zdeg(Q)\le l$. Theorem \ref{thmakwe} guarantees the list decoder with these parameters $m,l$ will correctly decode (that is, the list of roots contains the original message function) when there are at most $\lceil n-w/m\rceil-1$ errors.

\begin{exmp}[continued]
$G(0)=1,G(1)=0$, and $G(i)=1$ for $i\ge 2$ since $g=1$. Recalling that $u=4$, we have
\[
	\begin{array}{c|*{16}{c|}c}
		i   & 0 & 1 & 2 & 3 &4 &5 &6 &7 &8 &9 &10 &11 &12 & 13&14 &15 &\cdots\\\hline
	C(i)& 1 & 0 & 1 & 1 &2 &1 &2 &2 &3 &2 &3  &3  &4  &3  &4  &4 &\cdots\\\hline
	\sum_{j=0}^i C(j)	& 1 & 1 & 2 & 3 &5 &6 &8 &10&13&15&18&21&25&28&32&36&\cdots
	\end{array}
\]
For $m=2$, $N=25$. So $w=12$, and $l=3$. By the argument above, the list decoder with parameters $m=2,l=3$ is guaranteed to decode one arbitrary error.

Using the same bounds, successful decoding for two arbitrary errors is guaranteed if we take parameters $m=6,l=8$. Thus the successful decoding of two errors in the example with parameters $m=2,l=2$ is not to be expected from the bounds we have. In fact, our experiments show that decoding failures for two errors with parameters $m=2,l=2$ are actually infrequent. We expect that the bounds given above significantly underestimate the capability of the algorithm.
\end{exmp}

\section{Concluding Remarks}

We formulated list decoding of Hermitian codes anew, and presented a simple and efficient algorithm for the interpolation step. It is not easy to compare fairly our interpolation algorithm with previously known algorithms \cite{shokrollahi1999,hoeholdt1999,sakata2001,olshevsky2003}. However, our algorithm has a good computational complexity while its simple description affords a straightford hardware implementation.

The interpolation algorithm for Reed-Solomon codes in \cite{kwankyu2006} was shown to be equivalent to the Berlekamp-Massey algorithm in the special case when the multiplicity and list size parameters are all one. We expect that there is also an intimate relation between our interpolation algorithm for Hermitian codes with multiplicity and list size parameters all one and K\"otter's version of the Berlekamp-Massey-Sakata algorithm \cite{koetter1998,osullivan2004}.

Present bounds for the $Q$-polynomial need to be improved. In experiments, our list decoder with certain multiplicity and list size parameters shows a better rate of successful decoding than would be expected from the present bounds. A better understanding of the capability of the list decoder is required.

Though we try to make our formulation of list decoding as independent as possible from special properties of Hermitian codes, it is not clear what is the most general class of algebraic geometry codes for which list decoding is possible in a similar fashion.

\appendix

\section{A Gr\"obner Basis Algorithm}

We consider a submodule $S$ of $k[x]^m$. Let $e_1<e_2<\cdots<e_m$ denote the standard basis of $k[x]^m$. Let $u=(u_x,u_1,u_2,\dots,u_m)$ be a given sequence of positive integers. The $u$-weighted degree of a monomial $x^re_i$ is defined to be $\deg_u(x^re_i)=u_xr+u_i$. Thus $\deg_u(ae_i)=u_x\deg(a)+u_i$ for $a\in k[x]$. A monomial order $>_u$ of $k[x]^m$ is defined by declaring $x^re_i>_ux^se_j$ if $\deg_u(x^re_i)>\deg_u(x^se_j)$ or if $i>j$ when the weighted degrees are tied.

For $f=\sum_{i=1}^ma_ie_i$ with $a_i\in k[x]$, define the index of $f$, written $\ind(f)$, to be the largest $i$ such that $a_i\neq 0$. In particular, $\ind(x^re_i)=i$.

Suppose $\set{G_1,G_2,\dots,G_m}$ is a set of generators of the module $S$ such that $\ind(G_i)=i$. Then the following algorithm computes a Gr\"obner basis of $S$ from the given set of generators with respect to the monomial order $>_u$.

\begin{algG} 
Let $g_i=\sum_{j=1}^ma_{ij}e_j$ for $1\le i\le m$ during the execution of the algorithm. Initialize with $g_i\leftarrow G_i$.

\begin{enumerate}
\item[G1.] Set $r\leftarrow 1$.
\item[G2.] Increase $r$ by $1$. If $r\le m$, then proceed; otherwise go to step G6.
\item[G3.] Set $s\leftarrow \ind(\LT(g_r))$. If $s=r$, then go to step G2.
\item[G4.] Set $d\leftarrow\deg(a_{rs})-\deg(a_{ss})$ and  $c\leftarrow\LC(a_{rs})\LC(a_{ss})^{-1}$.
\item[G5.] If $d\ge 0$, then set 
\[
	g_r\leftarrow g_r-cx^dg_s.
\]
If $d<0$, then set, storing $g_s$ in a temporary variable,
\[
	g_s\leftarrow g_r,\quad g_r\leftarrow x^{-d}g_r-cg_s.
\]
Go back to step G3.
\item[G6.] Output $\set{g_1,\dots,g_m}$ and the algorithm terminates.  
\end{enumerate}
\end{algG}

\begin{prop}\label{prophfq} 
For each $1\le r\le m$, it occurs that $\ind(\LT(g_i))=i$ for $1\le i\le r$ after a finite number of iterations through the steps G3--G5.
\end{prop}

\begin{proof}
We actually prove that the following hold after initialization and after the iteration steps G3--G5:
\begin{itemize}
\item[(i)] $\ind(g_i)\le r$ for $1\le i\le r$.
\item[(ii)] $\ind(\LT(g_i))=i$ for $1\le i\le r-1$.
\item[(iii)] for every non-identity permutation $\pi$ of $\set{1,2,\dots,r}$,
\[
	\sum_{i=1}^r\deg_u(a_{ii}e_i)>\sum_{i=1}^r\deg_u(a_{i\pi_i}e_{\pi_i}).
\]
\end{itemize}
After initialization in step G1, when $r=1$, items (i)--(iii) are true. After $r$ is increased by one in step G2, (i) and (ii) clearly hold; (iii) also holds for the case $\pi_r=r$ because it holds for the previous value of $r$, and for the case $\pi_r\neq r$ because $\deg_u(a_{i\pi_i}e_{\pi_i})=-\infty$ for the $i$ such that $\pi_i=r$. It remains to check (i)--(iii) after the update in step G5. Item (i) is clear. Item (ii) still holds because leading terms of both $g_s$ and $g_r$ have index $s$. Explicitly
\begin{gather}
\label{equdjd}
	\deg_u(a_{sj}e_j)<\deg_u(a_{ss}e_s)\ge \deg_u(a_{s{j'}}e_{j'}),\\
\label{equdjs}
	\deg_u(a_{rj}e_j)<\deg_u(a_{rs}e_s)\ge \deg_u(a_{r{j'}}e_{j'})
\end{gather}
for $r\ge j>s>j'\ge 0$. Using \eqref{equdjd}, \eqref{equdjs}, and (iii), we can prove Propositions \ref{propdjca} and \ref{propqqqw}. Then using A1, B1, (ii), and (iii), we can prove Propositions~\ref{propcjwz} and \ref{propcefd}, which show that (iii) still holds after the update in step G5. 

Finally Propositions \ref{propdjca} and \ref{propqqqw} show that after the update in step G5, either $\deg_u(\LT(g_r))-\deg_u(a_{rr}e_r)$ strictly decreases or else the index of $\LT(g_r)$ strictly decreases. Therefore it will eventually happen that $\ind(\LT(g_r))=r$, which together with (ii) completes the proof.
\end{proof}

\begin{prop}\label{propdjca}
Assume the case $d\ge 0$. Let 
\[
	g=g_r-cx^dg_s=b_{rr}e_r+\cdots+b_{rs}e_s+\cdots,
\]
where $b_{rj}=a_{rj}-cx^da_{sj}$ for $0\le j\le r$. Then the following hold.
\begin{enumerate}
\item[\normalfont A1.] $\deg_u(b_{rr}e_r)=\deg_u(a_{rr}e_r)$.
\item[\normalfont A2.] $\deg_u(b_{rs}e_s)<\deg_u(a_{rs}e_s)$.
\item[\normalfont A3.] $\deg_u(b_{rj}e_j)<\deg_u(a_{rs}e_s)$ for $s<j<r$.
\item[\normalfont A4.] $\deg_u(b_{rj}e_j)\le\deg_u(a_{rs}e_s)$ for $j<s$.
\end{enumerate}
In particular, 
\[
	\deg_u(\LT(g))-\deg_u(b_{rr}e_r)\le\deg_u(\LT(g_r))-\deg_u(a_{rr}e_r),
\]
where the equality holds only if $\ind(\LT(g))<\ind(\LT(g_r))$.
\end{prop}

\begin{proof}
From (iii), choosing for $\pi$ the transposition of $s$ and $r$, we have $\deg(a_{rr})+\deg(a_{ss})>\deg(a_{rs})+\deg(a_{sr})$. Then A1 follows since
\[
	\deg(x^da_{sr})=\deg(a_{rs})-\deg(a_{ss})+\deg(a_{sr})<\deg(a_{rr}).
\]
A2 holds since $c$ and $d$ were chosen such that
\[
	\deg(b_{rs})=\deg(a_{rs}-cx^da_{ss})<\deg(a_{rs}).
\]
Let $s<j<r$. By \eqref{equdjd},
\[
	\deg_u(x^da_{sj}e_j)=\deg_u(a_{rs}e_s)-\deg_u(a_{ss}e_s)+\deg_u(a_{sj}e_j)<\deg_u(a_{rs}e_s),
\]
which together with \eqref{equdjs} shows A3.
Let $j<s$. Similarly by \eqref{equdjd},
\[
\deg_u(x^da_{sj}e_j)=\deg_u(a_{rs}e_s)-\deg_u(a_{ss}e_s)+\deg_u(a_{sj}e_j)\le\deg_u(a_{rs}e_s),
\]
which together with \eqref{equdjs} shows A4.

Now we show that the last assertion follows from A1--A4. Let $j=\ind(\LT(g))$ so that $\LT(g)=\LT(b_{rj}e_j)$. Recall that $\LT(g_r)=\LT(a_{rs}e_s)$. If $j=r$, then the assertion is obvious. Suppose $j<r$. Then by A2--A4, 
\[
	\deg_u(\LT(g))\le\deg_u(\LT(g_r)),
\]
where the equality holds only if $\ind(\LT(g))<\ind(\LT(g_r))$. On the other hand, $\deg_u(b_{rr}e_r)=\deg_u(a_{rr}e_r)$ by A1. The assertion follows.
\end{proof}

\begin{prop}\label{propcjwz}
Assume $d\ge0$, and retain previous notations. For every non-identity permutation $\pi=(\pi_1,\pi_2,\dots,\pi_r)$ of $\set{1,2,\dots,r}$,
\begin{equation}\label{equcjwq}
	\sum_{i\neq r}\deg_u(a_{ii}e_i)+\deg_u(b_{rr}e_r)>\sum_{i\neq r}\deg_u(a_{i\pi_i}e_{\pi_i})+\deg_u(b_{r\pi_r}e_{\pi_r}).
\end{equation}
\end{prop}

\begin{proof}
By A1, the left hand side of \eqref{equcjwq} equals 
\[
	\sum_{i=1}^r\deg_u(a_{ii}e_i).
\]
Recall that $b_{r\pi_r}=a_{r\pi_r}-cx^da_{s\pi_r}$. Hence $\deg(b_{r\pi_r})\le\deg(a_{r\pi_r})$ or $\deg(b_{r\pi_r})=\deg(a_{rs})-\deg(a_{ss})+\deg(a_{s\pi_r})$. 

If $\deg(b_{r\pi_r})\le\deg(a_{r\pi_r})$, then 
the right hand side of \eqref{equcjwq} is 
\[
	\le\sum_{i\neq r}\deg_u(a_{i\pi_i}e_{\pi_i})+\deg_u(a_{r\pi_r}e_{\pi_r})=\sum_i\deg_u(a_{i\pi_i}e_{\pi_i})<\sum_i\deg_u(a_{ii}e_i),
\]
where the last inequality holds by (iii). Thus \eqref{equcjwq} holds. 

Now we consider the case when $\deg(b_{r\pi_r})=\deg(a_{rs})-\deg(a_{ss})+\deg(a_{s\pi_r})$. Let $D_{ij}$ denote $\deg_u(a_{ij}e_j)$. Then \eqref{equcjwq} is equivalent to 
\begin{equation}\label{eqppwk}
	\sum_{i}D_{ii}>\sum_{i\neq s,r}D_{i\pi_i}+D_{s\pi_s}+D_{s\pi_r}+D_{rs}-D_{ss}.
\end{equation}
To show \eqref{eqppwk}, we treat two cases depending on whether $s$ and $\pi_r$ are in the same orbit or not, with respect to the permutation $\pi$.	First suppose $s$ and $\pi_r$ are in the same orbit so that
\[
\pi_r\stackrel{\pi}{\longrightarrow}\pi(\pi_r)\stackrel{\pi}{\longrightarrow}\dots\stackrel{\pi}{\longrightarrow}\pi^{-1}(s)\stackrel{\pi}{\longrightarrow} s.
\]
Let $S=\set{\pi_r,\pi(\pi_r),\dots,\pi^{-1}(s)}$. Note that $S$ is empty if $\pi_r=s$. Now the right hand side of \eqref{eqppwk} equals
\begin{equation}\label{equcnxx}
\begin{split}
	&\sum_{i\in S}D_{i\pi_i}+\sum_{i\not\in S,\, i\neq s,r}\!\!\!D_{i\pi_i}+D_{s\pi_s}+D_{rs}-D_{ss}+D_{s\pi_r}\\
	&\quad\le\sum_{i\in S}D_{ii}+\sum_{i\not\in S,\, i\neq s,r}\!\!\!D_{i\pi_i}+D_{s\pi_s}+D_{rs}.
\end{split}
\end{equation}
This inequality holds since $D_{ii}\ge D_{ij}$ for $1\le i\le r-1$ and $1\le j\le r$ by (ii). 
We can check that the second indices of the terms in the last expression of \eqref{equcnxx} are all distinct. So by (iii), we have
\[
	\sum_{i\in S}D_{ii}+\sum_{i\not\in S,\, i\neq s,r}\!\!\!D_{i\pi_i}+D_{s\pi_s}+D_{rs}<\sum_{i}D_{ii}.
\]
Hence \eqref{eqppwk} is proved.

The diagram (a) in Figure \ref{fig1} below gives an example exhibiting the intuition behind the argument above, for the case when $s$ and $\pi_r$ are in the same orbit. In the diagram, the smaller circles mark the terms in the first sum of \eqref{equcnxx} and the larger circles mark those of the second sum. Similar diagrams will be helpful later.

\begin{figure}[h]
\begin{center}
\begin{minipage}[t]{.4\textwidth}
\includegraphics{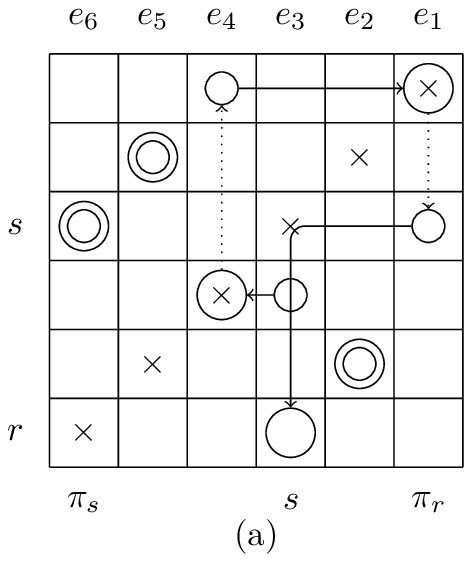}
\end{minipage}
\hfill
\begin{minipage}[t]{.5\textwidth}
\includegraphics{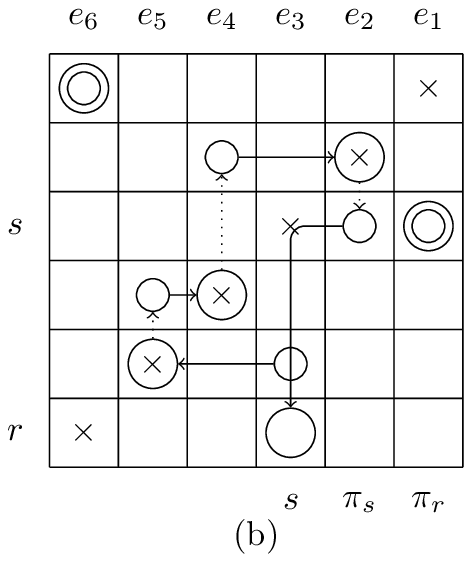}
\end{minipage}
\end{center}
\caption{\label{fig1}}
\end{figure}

If $s$ and $\pi_r$ are not in the same orbit, then we have
\[
\pi_s\stackrel{\pi}{\longrightarrow}\pi(\pi_s)\stackrel{\pi}{\longrightarrow}\dots\stackrel{\pi}{\longrightarrow}\pi^{-1}(s)\stackrel{\pi}{\longrightarrow} s,
\]
and let $S=\set{\pi_s,\pi(\pi_s),\dots,\pi^{-1}(s)}$. Note that $S$ is empty if $\pi_s=s$. Now the right hand side of \eqref{eqppwk} equals
\[
\begin{split}
	&\sum_{i\in S}D_{i\pi_i}+\sum_{i\not\in S,\, i\neq s,r}\!\!\!D_{i\pi_i}+D_{s\pi_r}+D_{rs}-D_{ss}+D_{s\pi_s}\\
	&\quad\le\sum_{i\in S}D_{ii}+\sum_{i\not\in S,\, i\neq s,r}\!\!\!D_{i\pi_i}+D_{s\pi_r}+D_{rs}<\sum_{i}D_{ii},
\end{split}
\]
where the inequalities are justified by similar arguments as above. See the diagram (b) in Figure \ref{fig1}.
\end{proof}

\begin{prop}\label{propqqqw}
Assume the case $d<0$. Let 
\[
	g=x^{-d}g_r-cg_s=b_{rr}e_r+\dots+b_{rs}e_s+\cdots,
\]
where $b_{rj}=x^{-d}a_{rj}-ca_{sj}$ for $0\le j\le r$. Then the following hold
\begin{enumerate}
\item[\normalfont B1.] $\deg_u(b_{rr}e_r)=\deg_u(x^{-d}a_{rr}e_r)$.
\item[\normalfont B2.] $\deg_u(b_{rs}e_s)<\deg_u(x^{-d}a_{rs}e_s)$.
\item[\normalfont B3.] $\deg_u(b_{rj}e_j)<\deg_u(x^{-d}a_{rs}e_s)$ for $s<j<r$.
\item[\normalfont B4.] $\deg_u(b_{rj}e_j)\le\deg_u(x^{-d}a_{rs}e_s)$ for $j<s$.
\end{enumerate}
In particular, 
\[
	\deg_u(\LT(g))-\deg_u(b_{rr}e_r)\le\deg_u(\LT(g_r))-\deg_u(a_{rr}e_r),
\]
where the equality holds only if $\ind(\LT(g))<\ind(\LT(g_r))$.
\end{prop}

\begin{proof}
B1 holds since
\[
	\deg(x^{-d}a_{rr})=\deg(a_{ss})-\deg(a_{rs})+\deg(a_{rr})>\deg(a_{sr})
\]
by (iii). B2 holds because $c$ and $d$ were chosen such that 
\[
	\deg(b_{rs})=\deg(x^{-d}a_{rs}-ca_{ss})<\deg(x^{-d}a_{rs}).
\] 
Let $s<j<r$. By \eqref{equdjd} and \eqref{equdjs},
\begin{gather*}
	\deg_u(x^{-d}a_{rj}e_j)<\deg_u(x^{-d}a_{rs}e_s),\\
	\deg_u(a_{sj}e_j)<\deg_u(a_{ss}e_s)=\deg_u(x^{-d}a_{rs}e_s),
\end{gather*}
from which B3 follows.
Let $j<s$. Again by \eqref{equdjd} and \eqref{equdjs},
\begin{gather*}
	\deg_u(x^{-d}a_{rj}e_j)\le\deg_u(x^{-d}a_{rs}e_s),\\
	\deg_u(a_{sj}e_j)\le\deg_u(a_{ss}e_s)=\deg_u(x^{-d}a_{rs}e_s),
\end{gather*}
from which B4 follows.

Now we show that the last assertion follows from B1--B4. Let $j=\ind(\LT(g))$ so that $\LT(g)=\LT(b_{rj}e_j)$. Recall that $\LT(g_r)=\LT(a_{rs}e_s)$. If $j=r$, then the assertion is obvious. Suppose $j<r$. Then by B2--B4, 
\[
	\deg_u(\LT(g))\le\deg_u(x^{-d}\LT(g_r)),
\]
where the equality holds only if $\ind(\LT(g))<\ind(\LT(g_r))$. On the other hand, $\deg_u(b_{rr}e_r)=\deg_u(x^{-d}a_{rr}e_r)$ by B1. The assertion follows.
\end{proof}

\begin{prop}\label{propcefd}
Assume the case $d<0$, and retain previous notations. For every non-identity permutation $\pi=(\pi_1,\pi_2,\dots,\pi_r)$ of $\set{1,2,\dots,r}$,
\begin{multline}\label{equhfgc}
	\sum_{i\neq s,r}\deg_u(a_{ii}e_i)+\deg_u(a_{rs}e_s)+\deg_u(b_{rr}e_r)\\
	>\sum_{i\neq s,r}\deg_u(a_{i\pi_i}e_{\pi_i})+\deg_u(a_{r\pi_s}e_{\pi_s})+\deg_u(b_{r\pi_r}e_{\pi_r}).
\end{multline}
\end{prop}

\begin{proof}
By B1, $\deg_u(b_{rr}e_r)=u_x(\deg(a_{ss})-\deg(a_{rs}))+\deg_u(a_{rr}e_r)$. Therefore the left hand side of \eqref{equhfgc} equals
\[
	\sum_{i=1}^r\deg_u(a_{ii}e_i).
\]
Note that $b_{r\pi_r}=x^{-d}a_{r\pi_r}-ca_{s\pi_r}$. Hence $\deg(b_{r\pi_r})\le\deg(a_{s\pi_r})$ or $\deg(b_{r\pi_r})=\deg(a_{ss})-\deg(a_{rs})+\deg(a_{r\pi_r})$. 

If $\deg(b_{r\pi_r})\le\deg(a_{s\pi_r})$, then 
the right hand side of \eqref{equhfgc} is 
\[
	\le\sum_{i\neq s,r}\deg_u(a_{i\pi_i}e_{\pi_i})+\deg_u(a_{r\pi_s}e_{\pi_s})+\deg_u(a_{s\pi_r}e_{\pi_r})<\sum_i\deg_u(a_{ii}e_i),
\]
where the last inequality holds by (iii). Thus \eqref{equhfgc} holds. 

Now we suppose $\deg(b_{r\pi_r})=\deg(a_{ss})-\deg(a_{rs})+\deg(a_{r\pi_r})$. Let $D_{ij}$ denote $\deg_u(a_{ij}e_j)$. Note that \eqref{equhfgc} is equivalent to 
\begin{equation}\label{equchqw}
	\sum_{i}D_{ii}>\sum_{i\neq s,r}D_{i\pi_i}+D_{r\pi_s}+D_{r\pi_r}+D_{ss}-D_{rs}.
\end{equation}
To show this, we treat two cases depending on whether $s$ and $\pi_r$ are in the same orbit or not, with respect to the permutation $\pi$.	First suppose $s$ and $\pi_r$ are in the same orbit so that
\[
\pi_r\stackrel{\pi}{\longrightarrow}\pi(\pi_r)\stackrel{\pi}{\longrightarrow}\dots\stackrel{\pi}{\longrightarrow}\pi^{-1}(s)\stackrel{\pi}{\longrightarrow} s.
\]
Let $S=\set{\pi_r,\pi(\pi_r),\dots,\pi^{-1}(s)}$. Note that $S$ is empty if $\pi_r=s$. Now the right hand side of \eqref{equchqw} equals
\begin{equation}\label{equcnvf}
\begin{split}
	&\sum_{i\in S}D_{i\pi_i}+\sum_{i\not\in S,\, i\neq s,r}\!\!\!D_{i\pi_i}+D_{r\pi_s}+D_{ss}-D_{rs}+D_{r\pi_r}\\
	&\quad\le\sum_{i\in S}D_{ii}+\sum_{i\not\in S,\, i\neq s,r}\!\!\!D_{i\pi_i}+D_{r\pi_s}+D_{ss}.
\end{split}
\end{equation}
This inequality holds since $D_{ii}\ge D_{ij}$ for $1\le i\le r-1$ and $1\le j\le r$ by (ii) and that $D_{rs}\ge D_{rj}$ for $1\le j\le r$ by the way in which $s$ is chosen. We can check that the right indices of terms in the final expression of \eqref{equcnvf} are all distinct. So by (iii), we see that
\[
	\sum_{i\in S}D_{ii}+\sum_{i\not\in S,\, i\neq s,r}\!\!\!D_{i\pi_i}+D_{r\pi_s}+D_{ss}<\sum_{i}D_{ii}.
\]
Hence \eqref{equchqw} is proved. See the diagram (c) in Figure \ref{fig2}.

If $s$ and $\pi_r$ are not in the same orbit, then we have
\[
\pi_s\stackrel{\pi}{\longrightarrow}\pi(\pi_s)\stackrel{\pi}{\longrightarrow}\dots\stackrel{\pi}{\longrightarrow}\pi^{-1}(s)\stackrel{\pi}{\longrightarrow} s,
\]
and let $S=\set{\pi_s,\pi(\pi_s),\dots,\pi^{-1}(s)}$. Note that $S$ is empty if $\pi_s=s$. Now the right hand side of \eqref{equchqw} equals
\[
\begin{split}
	&\sum_{i\in S}D_{i\pi_i}+\sum_{i\not\in S,\, i\neq s,r}\!\!\!D_{i\pi_i}+D_{r\pi_r}+D_{ss}-D_{rs}+D_{r\pi_s}\\
	&\quad\le\sum_{i\in S}D_{ii}+\sum_{i\not\in S,\, i\neq s,r}\!\!\!D_{i\pi_i}+D_{r\pi_r}+D_{ss}<\sum_{i}D_{ii},
\end{split}
\]
where the inequalities hold by the same reasons as above. See the diagram (d) in Figure \ref{fig2}.

\begin{figure}[h]
\begin{center}
\begin{minipage}[t]{.4\textwidth}
\includegraphics{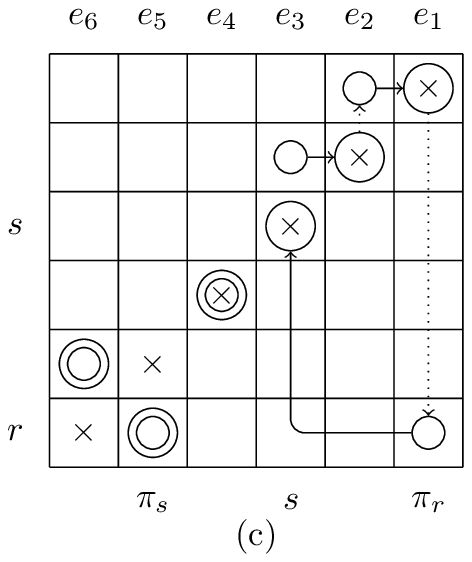}
\end{minipage}
\hfill
\begin{minipage}[t]{.5\textwidth}
\includegraphics{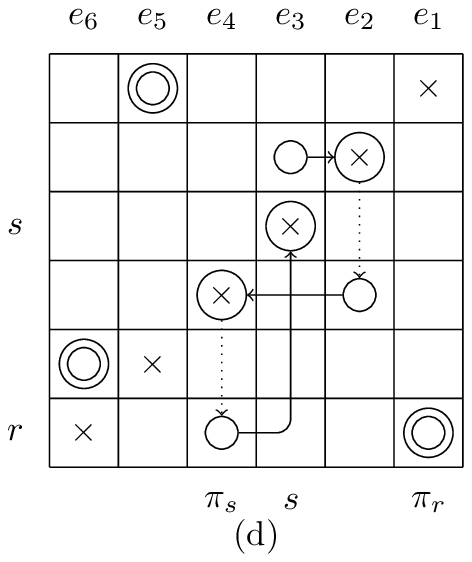}
\end{minipage}
\end{center}
\caption{\label{fig2}}
\end{figure}

\end{proof}

\begin{prop}\label{propxaaa}
Let $g_i=\sum_{j=1}^ma_{ij}e_j$, $1\le i\le m$ be an input for the algorithm. Let $c$ be an upper bound on $\deg_u(a_{ij}e_j)$, $1\le i,j\le m$. Let $d$ be an upper bound on $\deg_u(a_{ij}e_j)-\deg_u(a_{ii}e_i)$, $1\le j\le i\le m$. Then an execution of the algorithm for the input $g_i$ requres $O(cu_x^{-1}dm^3)$ multiplication operations in the field $k$.
\end{prop}

\begin{proof}
The proof of Proposition \ref{prophfq} implies that for each $1\le i\le m$, at most $i(\deg_u(\LT(g_i))-\deg_u(a_{ii}e_i))$ updates for $g_i$ are performed. Note that $d\ge\deg_u(\LT(g_i))-\deg_u(a_{ii}e_i)$. Note also that each update requires at most $cu_x^{-1}i$ multiplication operations in $k$. Therefore at most $\sum_{i=1}^mcu_x^{-1}di^2$ multiplication operations are required.
\end{proof}


\begin{thebibliography}{10}

\bibitem{augot2000}
D.~Augot and L.~Pecquet.
\newblock A {Hensel} lifting to replace factorization in list-decoding of
  algebraic-geometric and {Reed-Solomon} codes.
\newblock {\em {IEEE} Trans. Inform. Theory}, 46(7):2605--2614, 2000.

\bibitem{cox1997}
D.~Cox, J.~Little, and D.~O'Shea.
\newblock {\em Ideals, Varieties, and Algorithms}.
\newblock Springer-Verlag, New York, second edition, 1997.

\bibitem{cox1997a}
D.~Cox, J.~Little, and D.~O'Shea.
\newblock {\em Using Algebraic Geometry}, volume 185 of {\em GTM}.
\newblock Springer-Verlag, New York, 1998.

\bibitem{fulton1969}
W.~Fulton.
\newblock {\em Algebraic Curves}.
\newblock Benjamin, 1969.

\bibitem{gao2000}
S.~Gao and M.~A. Shokrollahi.
\newblock Computing roots of polynomials over function fields of curves.
\newblock In {\em Coding Theory and Cryptography: From Enigma and
  Geheimschreiber to Quantum Theory}, pages 214--228. Springer-Verlag, 2000.

\bibitem{guruswami1999}
V.~Guruswami and M.~Sudan.
\newblock Improved decoding of {Reed-Solomon} and algebraic-geometry codes.
\newblock {\em {IEEE} Trans. Inform. Theory}, 45(6):1757--1767, 1999.

\bibitem{hoeholdt1999}
T.~H{\o}holdt and R.~R. Nielsen.
\newblock Decoding {H}ermitian codes with {S}udan's algorithm.
\newblock In {\em Applied algebra, algebraic algorithms and error-correcting
  codes}, volume 1719 of {\em LNCS}, pages 260--270. Springer, 1999.

\bibitem{koetter1998}
R.~K\"otter.
\newblock A fast parallel implementation of a {Berlekamp-Massey} algorithm for
  algebraic-geometric codes.
\newblock {\em {IEEE} Trans. Inform. Theory}, 44(4):1353--1368, 1998.

\bibitem{kwankyu2006}
K.~Lee and M.~E. O'Sullivan.
\newblock Sudan's list decoding of {Reed-Solomon} codes from a {Gr\"obner}
  basis perspective.
\newblock 2006.
\newblock arXiv:math.AC/0601022.

\bibitem{olshevsky2003}
V.~Olshevsky and M.~A. Shokrollahi.
\newblock A displacement approach to decoding algebraic codes.
\newblock In {\em Fast algorithms for structured matrices: theory and
  applications}, volume 323 of {\em Contemp. Math.}, pages 265--292. Amer.
  Math. Soc., 2003.

\bibitem{osullivan2004}
M.~E. O'Sullivan.
\newblock On {Koetter's} algorithm and the computation of error values.
\newblock {\em Designs, Codes and Cryptography}, 31:169--188, 2004.

\bibitem{pretzel1998}
O.~Pretzel.
\newblock {\em Codes and Algebraic Curves}.
\newblock Oxford UP, 1998.

\bibitem{sakata2001}
S.~Sakata.
\newblock On fast interpolation method for {Guruswami-Sudan} list decoding of
  one-point algebraic-geometry codes.
\newblock In {\em Applied algebra, algebraic algorithms and error-correcting
  codes}, volume 2227 of {\em LNCS}, pages 172--181. Springer, 2001.

\bibitem{shokrollahi1999}
M.~A. Shokrollahi and H.~Wasserman.
\newblock List decoding of algebraic-geometric codes.
\newblock {\em {IEEE} Trans. Inform. Theory}, 45(2):432--437, 1999.

\bibitem{stichtenoth1993}
H.~Stichtenoth.
\newblock {\em Algebraic Function Fields and Codes}.
\newblock Springer-Verlag, 1993.

\bibitem{sudan1997}
M.~Sudan.
\newblock Decoding of {Reed-Solomon} codes beyond the error-correction bound.
\newblock {\em J. Complexity}, 13(1):180--193, 1997.

\bibitem{wu2001}
X.-W. Wu and P.~H. Siegel.
\newblock Efficient root-finding algorithm with applications to list decoding
  of algebraic-geometric codes.
\newblock {\em {IEEE} Trans. Inform. Theory}, 47(6):2579--2587, 2001.

\end{thebibliography}
\end{document}